\documentclass[runningheads]{llncs}
\usepackage{graphicx}
\usepackage{subfigure}
\usepackage[]{hyperref}
\usepackage{array}
\usepackage{amsfonts}
\usepackage{algorithm}
\usepackage{algorithmicx}
\usepackage{algpseudocode}
\usepackage{amsmath}
\usepackage{indentfirst}
\usepackage{booktabs}
\usepackage{multirow}
\usepackage[misc]{ifsym} 
\usepackage{color}
\begin{document}

\toctitle{EpiGNN: Exploring Spatial Transmission with Graph Neural Network for Regional Epidemic Forecasting} 
\tocauthor{Feng Xie, Zhong Zhang, Liang Li, Bin Zhou, Yusong Tan}

\title{EpiGNN: Exploring Spatial Transmission \\ with Graph Neural Network for Regional Epidemic Forecasting}
\titlerunning{EpiGNN: Exploring Spatial Transmission with GNN}
%
\author{Feng Xie \and
Zhong Zhang \and
Liang Li \and
Bin Zhou(\Letter) \and
Yusong Tan}
\authorrunning{F. Xie et al.}
%
\institute{College of Computer, National University of Defense Technology
\email{\{xiefeng,zhangzhong,liliang98,binzhou,ystan\}@nudt.edu.cn}}
\maketitle              

\begin{abstract}

\let\thefootnote\relax\footnotetext{\Letter {} Corresponding author.}
Epidemic forecasting is the key to effective control of epidemic transmission and helps the world mitigate the crisis that threatens public health. To better understand the transmission and evolution of epidemics, we propose EpiGNN, a graph neural network-based model for epidemic forecasting. Specifically, we design a transmission risk encoding module to characterize local and global spatial effects of regions in epidemic processes and incorporate them into the model. Meanwhile, we develop a Region-Aware Graph Learner (RAGL) that takes transmission risk, geographical dependencies, and temporal information into account to better explore spatial-temporal dependencies and makes regions aware of related regions' epidemic situations. The RAGL can also combine with external resources, such as human mobility, to further improve prediction performance. Comprehensive experiments on five real-world epidemic-related datasets (including influenza and COVID-19) demonstrate the effectiveness of our proposed method and show that EpiGNN outperforms state-of-the-art baselines by 9.48\% in RMSE.

\end{abstract}
\section{Introduction}

\noindent Epidemics spread through human-to-human interaction and circulate worldwide, seriously endangering public health. The World Health Organization (WHO) estimates that seasonal influenza annually causes approximately 3–5 million severe cases and 290,000–650,000 deaths.\footnote{https://www.who.int/en/news-room/fact-sheets/detail/influenza-(seasonal)} Recently, the coronavirus disease 2019 (COVID-19) has spread over more than 200 countries and territories,\footnote{https://covid19.who.int/} causing heavy human losses and economic burdens. Accurate prediction of epidemics is the key to effective control of epidemic transmission and plays an essential role in driving administrative decision-making, timely allocating healthcare resources, and helping with drug research.

A number of studies have investigated epidemic forecasting for decades, aiming to help the world mitigate the crisis that threatens public health. In statistics community, autoregressive (AR) models are widely used in epidemic forecasting \cite{wang2015dynamic,chakraborty2019forecasting}. In compartment models, the susceptible-infected-recovered (SIR) is the most basic one, many cumulative works in this category are based on its extensions \cite{aron1984seasonality,won2017early}. However, the above methods are limited in accuracy and generalization due to their oversimplified or fixed assumptions. Recently, deep learning has achieved tremendous success in many challenging tasks, and various deep learning-based epidemic prediction models \cite{wu2018deep,adhikari2019epideep,jung2021self} have been proposed, especially models based on emerging graph neural networks (GNNs) \cite{deng2020cola,panagopoulos2021transfer,wang2022causalgnn}. The core insight behind GNNs is to capture correlations between nodes and model the signal propagation of neighbor nodes. In regional epidemic prediction task, GNN-based approaches model the spread of epidemics by regarding regions as nodes and hidden correlations between regions as edges in a graph structure.

\begin{figure}
\centering
\includegraphics[width=0.9\textwidth]{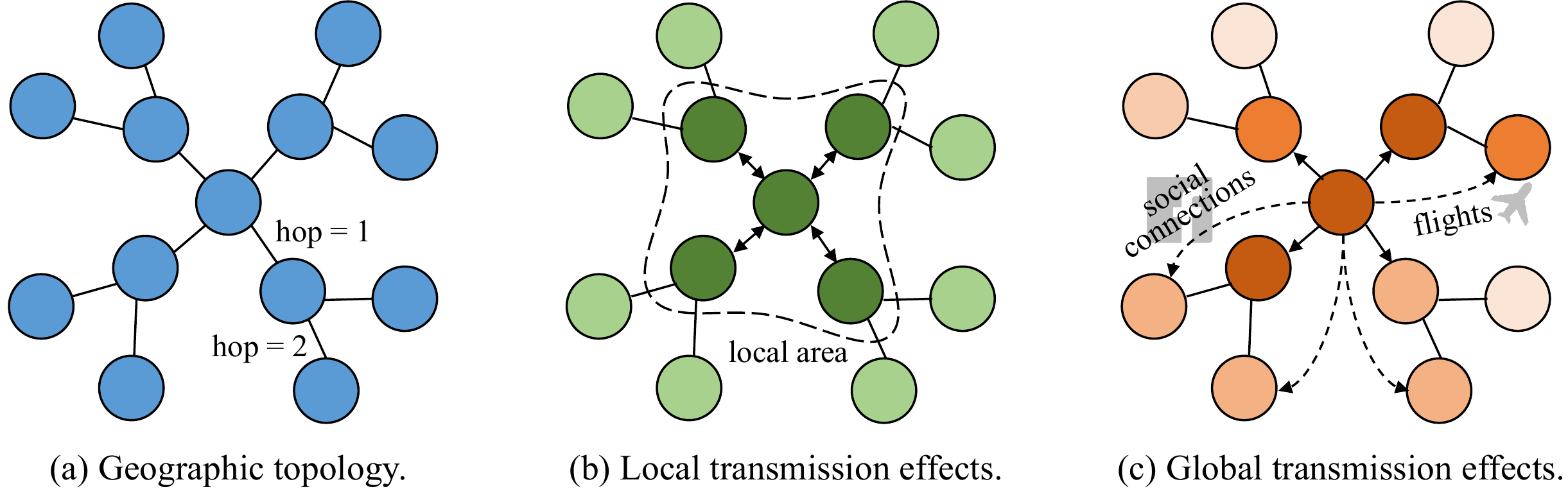}
\caption{The illustration of geographic topology, local and global spatial transmission effects, where nodes represent regions and edges represent the relationships.} \label{fig:spatialeffect}
\end{figure}

Although both spatial dependencies and temporal information are well exploited, existing methods still face two main challenges. First, the key to GNN-based models is to capture high-quality connections between regions. Using explicit graph structures, such as geographic topology (Fig. \ref{fig:spatialeffect}(a)), does not necessarily reflect the true dependencies or is hard to capture the hidden relationships \cite{wu2019graph}. Some effective works \cite{panagopoulos2021transfer,zhang2021multi} capture potential relationships between regions using specific data (e.g., human mobility) that require struggling with data availability, data accuracy, and data privacy. Due to the excellent feature extraction capability of the attention mechanism \cite{vaswani2017attention}, several studies \cite{deng2020cola,jung2021self} are mainly dedicated to combining attention mechanism and the latent representation of each region to capture correlations between regions based on similarity. However, owing to the global receptive field of attention mechanism, during aggregating features from other regions, it is prone to causing oversmoothing \cite{li2018deeper}, or bringing noise especially when the data is noisy and sparse in epidemic surveillance \cite{wang2022causalgnn}, which will damage the forecasting performance. Therefore, capturing underlying transmission dependencies between regions reasonably and accurately is crucial to facilitate further improving the prediction performance of GNN-based methods. At the same time, the method we expect should flexibly support both scenarios when rich external information can be collected or not.

Second, some studies \cite{han2021quantifying,mcmahon2022spatial} have paid much attention to mining the transmission factors of epidemics and assessing the spatial transmission risks of regions, and they suggest transmission risks are meaningful information that provides more practical insights for understanding the spread of epidemics. Spatial transmission risk implies \textsl{a potential ability that the epidemic in one region impacts other regions from a spatial perspective}, which is not only important property of regions but also reveals the spatial effects between regions. As shown in Fig. \ref{fig:spatialeffect}, in typical epidemic processes, virus tend to firstly spread in a local range due to intensive mobility of internal elements (e.g., human mobility) between geographically adjacent regions \cite{zhang2021multi} (Fig. \ref{fig:spatialeffect}(b)). Moreover, the epidemic in one region has not only local effects but also spillover effects across regions through complicated social connections \cite{han2021quantifying} (Fig. \ref{fig:spatialeffect}(c)). Thus, modeling spatial effects of regions are beneficial for understanding the spread and evolution of epidemics, which motivates us to investigate how to leverage regional transmission risk to enhance the accuracy and interpretability of epidemic prediction.

To tackle the aforementioned challenges and better understand the spread of epidemics, we propose a novel neural network model, termed EpiGNN, which handles temporal and spatial information through Convolution Neural Network and Graph Convolution Network. In this model, we propose a transmission risk encoding module to characterize spatial effects of regions. Meanwhile, we develop a Region-Aware Graph Learner which takes transmission risk, geographical information, and temporal dependencies into consideration to capture correlations between regions. Our contributions are summarized as follows:

\begin{itemize}
    \item We design a novel graph neural network-based model for epidemic prediction in which a transmission risk encoding module is proposed that shows how we incorporate local and global spatial effects of regions into the model.

    \item We introduce a Region-Aware Graph Learner which takes transmission risk, geographical information, and temporal dependencies into account to better explore underlying spatio-temporal correlations between regions.

    \item We evaluate our model on five epidemic-related datasets. Experimental results show the proposed method achieves state-of-the-art performance and demonstrate the effectiveness of our model. The source code and datasets are available at \url{https://github.com/Xiefeng69/EpiGNN}.
\end{itemize}

The remainder of this paper is organized as follows. We review related works in Section \ref{sec:RW}. Then we explain the details of our contributions in section \ref{sec:Method} and present experiments and results in Section \ref{sec:EXP}. At last, we conclude in Section \ref{sec:CON}.

\section{Related Work} \label{sec:RW}

\noindent \textbf{Epidemic forecasting methods.} As mentioned above, there has been a large body of work focusing on epidemic forecasting. Essentially, the aim of epidemic forecasting is to predict the number of infection cases for a region at a timestamp based on historical data. In statistics community, autoregressive (AR) models are widely used in epidemic forecasting \cite{wang2015dynamic,chakraborty2019forecasting}. In compartment models, susceptible-infected-recovered (SIR) is the most basic one that divides a population into three groups: susceptible, infected, and recovered, and simulates the variations over time between groups. Many cumulative works in this category are based on its extensions \cite{aron1984seasonality,won2017early}. Although these methods have a solid mathematical foundation, their accuracy and generalization are limited due to their oversimplified or fixed assumptions, pre-supposed functional form, and careful feature engineering. In recent years, due to its powerful data learning capability, deep learning has been widely adopted in various fields, including epidemic prediction tasks. Wu et al. \cite{wu2018deep} proposed CNNRNN-Res that firstly applied deep learning for epidemic forecasting. Adhikari et al. \cite{adhikari2019epideep} adopted deep clustering to help determine the historical season closest to the predicted time point to aid prediction. Jin et al. \cite{jin2021inter} introduced an inter-series attention-based model to capture similar progression patterns between time series to assist in COVID-19 prediction. Jung et al. \cite{jung2021self} designed a self-attention-based approach that cooperates with Long Short-Term Memory (LSTM) for regional influenza prediction. 

\textbf{Graph Neural Network-based models.} Graph neural networks (GNNs) have emerged in recent years, such as GCN \cite{kipf2016semi}, ST-GCN \cite{yu2017spatio}, and demonstrated promising results for extracting the correlation of irregular, non-Euclidean graph data, which make them become powerful tools for understanding the spread and evolution of epidemics. GNN-based epidemic prediction approaches create a graph where nodes correspond to regions of a country, and edge weights correspond to correlations between regions. Deng et al. \cite{deng2020cola} proposed Cola-GNN that applied an attention mechanism to learn the dependencies between regions based on the latent state of each region learned through Recurrent Neural Networks (RNNs). Panagopoulos et al. \cite{panagopoulos2021transfer} took advantage of mobility data across different regions to explore the underlying correlations between regions and adopted message passing neural network (MPNN) combined with LSTM to capture the spatial and temporal evolution of COVID-19. Zhang et al. \cite{zhang2021multi} developed a multi-modal information fusion-powered method that took social connections and demographic information into account to improve COVID-19 forecasting. Wang et al. \cite{wang2022causalgnn} designed CausalGNN which employed a causal module to provide epidemiological context for guiding the learning of spatial and temporal disease dynamics. Inspired by these works, we aim to explore spatial transmission in typical epidemic processes with GNNs for regional epidemic forecasting.



\section{The Proposed Method} \label{sec:Method}

\subsection{Problem Formulation}

\begin{figure}
\centering
\includegraphics[width=0.95\textwidth]{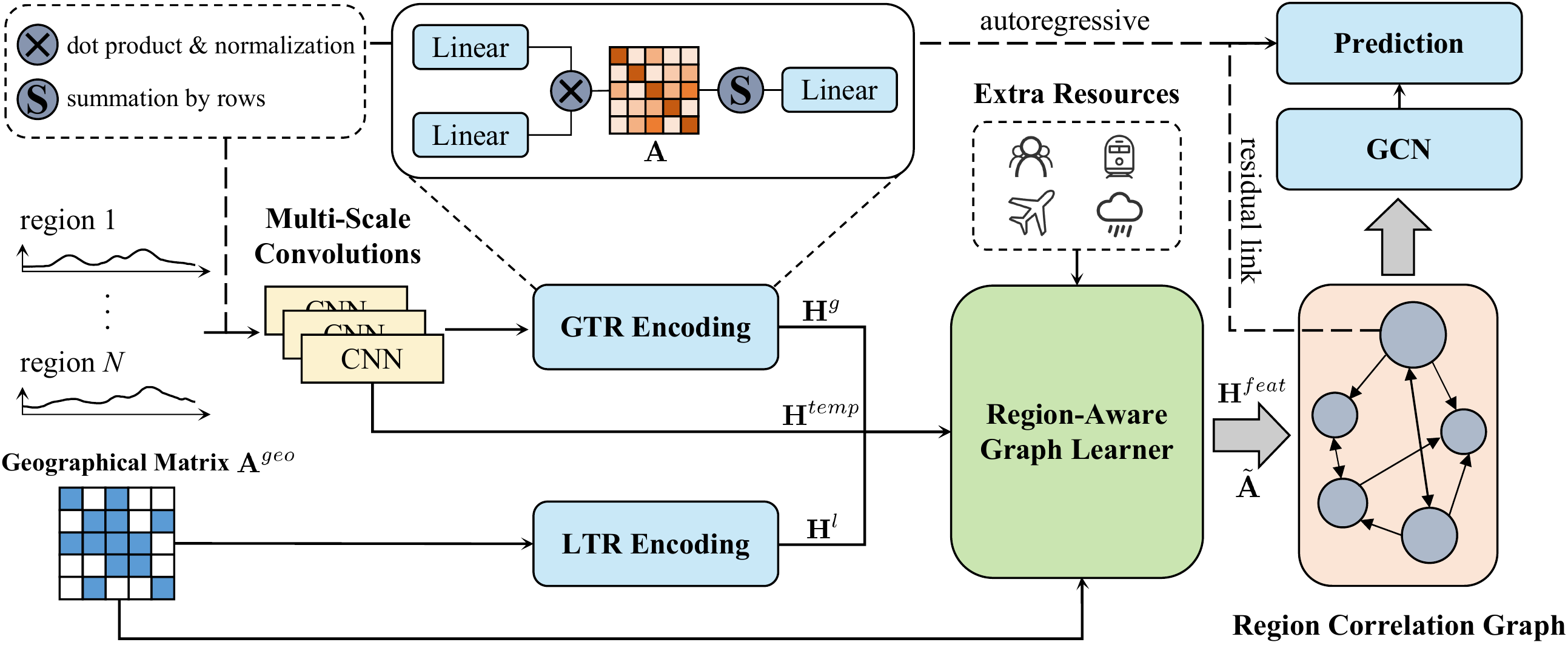}
\caption{The overview of our proposed method: EpiGNN.} \label{fig:epignn}
\end{figure}

\noindent We formulate the epidemic prediction problem as a graph-based propagation model. We have a total of $N$ regions (e.g., cities or states). We denote the historical cases data $\mathbf{X}=[\mathbf{x}_1,...,\mathbf{x}_{t}]$ as training data, where $\mathbf{x}_z\in\mathbb{R}^{N}$ represents the observed cases value of $N$ regions at time $z$. Our goal is to predict the future cases value, i.e. $\mathbf{x}_{t+h}$, where $h$ is a fixed horizon with respect to different tasks (e.g., short- or long-term prediction). For every task, we use $[\mathbf{x}_{t-T+1}...,\mathbf{x}_{t}]\in\mathbb{R}^{N\times{T}}$ for a look-back window $T$ to predict $\mathbf{x}_{t+h}$. For a region $i$, it is associated with a time series $\mathbf{x}_{i:}=[x_{i,t-T+1},...,x_{i,t}]$. The proposed method is drawn in Fig. \ref{fig:epignn}. In following sections, we introduce the building blocks for EpiGNN in detail.

\subsection{Multi-Scale Convolutions}

\noindent Convolutional Neural Networks (CNNs) have demonstrated strong feature representation ability and efficient parallel computation in grid data and sequence data that apply learnable filters to capture information behind data. Some works \cite{wu2020connecting,deng2020cola} suggest that using a set of multi-scale convolutions can capture complex temporal patterns simultaneously. Therefore, in this work, we also adopt multi-scale convolutions with different filter sizes and dilated factors as a feature extractor. We denote convolution filter as $\mathbf{f}_{1\times{s},d}$, where $s$ is filter size, $d$ is dilated factor, both $s$ and $d$ are empirically selected. The convolution operation of series $\mathbf{x}_{i:}$ with $\mathbf{f}_{1\times{s},d}$ at step $j$ is represented as:
\begin{equation}
    \mathbf{x}_{i:}\star{\mathbf{f}_{1\times{s},d}}(j)=\sum_{i=0}^{s-1}\mathbf{f}_{1\times{k}}(i)\mathbf{x}(j-d\times{i}),
\end{equation}

\noindent where $\star$ is convolution operator. We use $m$ parallel convolutional layers, each scale with $k$ filters, to generate different feature vectors, and concatenate them after an adaptive pooling layer. We denote $D=(m\times{k}\times{p})$ as the output dimension of multi-scale convolutions, where $p$ is the output dimension of adaptive pooling layer. At last, we obtain the temporal feature $\mathbf{h}_{i}^{temp}\in\mathbb{R}^{D}$ for region $i$.

\subsection{Transmission Risk Encoding Module}

\noindent The epidemic in one region has not only local effects but also spillover effects across regions through complicated social connections \cite{han2021quantifying}. Therefore, We assess local and global transmission risks for regions respectively and encode them as important properties of regions. Essentially, transmission risk encodings indicate spatial structure information which reflects potential spread influence of regions.

\subsubsection{Local Transmission Risk (LTR) Encoding.}

The proximity between regions will lead to a rapid increase in the mobility of internal elements between regions (e.g., human mobility), which will exacerbate local transmission risk. In the geographical network topology, the \textsl{degree} is a valuable signal for understanding network structure and describing the centrality of nodes. The more central regions will potentially interact with their surrounding regions more frequently, which leads to significant local spatial effects and is more likely to cause the virus to spread. Hence, we use the {degree} of each region in geographical topology to measure its local transmission risk. We generate local transmission risk encoding $\mathbf{h}_{i}^{l}\in\mathbb{R}^{D}$ by following equation:
\begin{equation}
  \mathbf{h}_{i}^{l} = \mathbf{W}^{l} \cdot {d}_{i} + \mathbf{b}^{l},
\end{equation}

\noindent where $d_{i}=\sum_{j}a^{geo}_{i,j}$ means the {degree} of region $i$, and $\mathbf{A}^{geo}$ is the geographical adjacency matrix that indicates the spatial connectivity of regions: $a^{geo}_{i,j}=1$ means region $i$ and region $j$ are neighbors (by default, $a^{geo}_{i,i}=1$). $\mathbf{W}^l$ and $\mathbf{b}^l$ are the parameters to transform degree vector $\mathbf{d}\in\mathbb{R}^{N}$ to encodings.

\subsubsection{Global Transmission Risk (GTR) Encoding.}
Besides geographical adjacent, there are also potential correlations between disjoint regions (e.g., social connections). During the spread of epidemics, it is highly likely that similar progression patterns are shared among related regions because they suffer from the same virus. We believe that if a region has a similar progression pattern to another region, there is probably a dependency between them. Therefore, for global transmission risk assessment, we measure it by the sum of the dynamic correlations based on temporal features of regions, and we call it the \textsl{global correlation coefficient} in this paper. Inspired by the self-attention \cite{vaswani2017attention}, we obtain global correlation coefficients and GTR encodings by following equations:
\begin{equation}
  \mathbf{A}=(\mathbf{H}^{temp}\mathbf{W}^{q})(\mathbf{H}^{temp}\mathbf{W}^{k})^{T},
\end{equation}
\begin{equation}
  a_{i,j}=\frac{a_{i,j}}{\text{max}(\left\| \mathbf{a}_{i:} \right\|_2,\epsilon)},
\end{equation}
\begin{equation}
  {g}_i=\sum_{j}a_{i,j},
\end{equation}
\begin{equation}
  \mathbf{h}_{i}^{g}= \mathbf{W}^{g} \cdot {g}_{i} + \mathbf{b}^{g},
\end{equation}

\noindent where $\mathbf{W}^q$, $\mathbf{W}^k\in\mathbb{R}^{D\times{F}}$, and $\mathbf{W}^g\in\mathbb{R}^{D}$ are weight matrices, $\epsilon$ is a small value to avoid division by zero. More precisely, we first feed temporal features $\mathbf{H}^{temp}$ to two parallel dense layers and apply a dot product to obtain a correlation distribution matrix $\mathbf{A}$. Then we adopt normalization for each row in $\mathbf{A}$ and calculate global correlation coefficient vector $\mathbf{g}\in\mathbb{R}^{N}$. At last, we feed ${g_{i}}$ to a dense layer to form global transmission risk encoding $\mathbf{h}_{i}^{g}\in\mathbb{R}^{D}$.

\subsection{Region-Aware Graph Learner}

\noindent Capturing correlations between regions by simulating all factors related to the spread of epidemics is troublesome, so we design a Region-Aware Graph Learner (RAGL), which considers both temporal and spatial information to generate a region correlation graph, where nodes correspond to regions, and edge weights correspond to the correlations between regions. We fuse temporal features and transmission risk encodings as nodes' initial attributes $\mathbf{H}^{feat}\in\mathbb{R}^{N\times{D}}$:

\begin{equation}
    \mathbf{h}_{i}^{feat}=\mathbf{h}_{i}^{temp}+\mathbf{h}_{i}^{l}+\mathbf{h}_{i}^{g}.
\end{equation}

Existing methods for learning correlations based on attention mechanisms are often symmetric or bidirectional \cite{wu2020connecting}. However, the epidemic transmission is often spread from one region to another, or one region impacts another, so we expect that the learned region correlation graph should not be a completely bidirectional graph. First, we extract dynamic temporal relationships by following equations:
\begin{equation}
  \mathbf{M}_1=\text{tanh}(\mathbf{H}^{temp}\mathbf{W}_{1}+\mathbf{b}_{1}),\quad  \mathbf{M}_2=\text{tanh}(\mathbf{H}^{temp}\mathbf{W}_{2}+\mathbf{b}_{2}),
\end{equation}
\begin{equation}
  \hat{\mathbf{A}}=\text{ReLU}(\text{tanh}(\mathbf{M}_{1}\mathbf{M}_{2}^{T}-\mathbf{M}_{2}\mathbf{M}_{1}^{T})),
\end{equation}

\noindent where $\mathbf{W}_1$, ${\mathbf{W}_2}\in\mathbb{R}^{D\times{F}}$ are weight matrices. The subtraction term and $\text{ReLU}(\cdot)$ regularize the connectivity of temporal correlation matrix $\hat{\mathbf{A}}$. Next, we capture spatial dependencies utilizing $\mathbf{A}^{geo}$, where we also introduce the \textsl{degree} to assess local spatial effects. Specifically, we use the product of the degrees of two adjacent regions as a gate that measures the impacts of local interactions between regions to control spatial dependencies:
\begin{equation}
  \mathbf{D}^{s}=\text{sigmoid}(\mathbf{W}^{s}\circ \mathbf{d}\mathbf{d}^{T}),
\end{equation}
\begin{equation}
  \tilde{\mathbf{A}}=\mathbf{D}^{s} \circ \mathbf{A}^{geo} + \hat{\mathbf{A}},
\end{equation}
\noindent where $\circ$ is element-wise (Hadamard) product, and $\mathbf{W}^{s}\in\mathbb{R}^{N\times{N}}$ is a learnable parameter matrix. The spread of epidemics is associated with many factors (e.g., human mobility, climate). RAGL can flexibly take advantage of external resources that are available to extract dependencies between regions more accurately. We denote external resources as $\mathbf{E}=[\mathbf{E}_1,\mathbf{E}_2,...,\mathbf{E}_{t}]$ where $\mathbf{E}_{z}\in\mathbb{R}^{N\times{N}}$ represents external correlation between regions at time step ${z}$ (e.g., the weight of edge $e^{z}_{i,j}$ represents the total number of people that moved from region $i$ to region $j$), we can calculate external correlation matrix by following equation:
\begin{equation}\label{eq:external}
    \mathbf{A}^{e}=\mathbf{W}^{e} \circ \sum_{i=0}^{e-1}\mathbf{E}_{t-e},
\end{equation}

\noindent where $e$ is the look-back window of external resources, and $\mathbf{W}^e$ is a learnable matrix. At last, we sum them up to obtain the region correlation matrix $\tilde{\mathbf{A}}$. 

\subsection{Graph Convolution Network}

\noindent Graph Convolution Networks (GCNs) as a kind of GNNs have been proven to be effective methods for learning node representations. In this work, we apply GCN to investigate the epidemic propagation among different regions \cite{kipf2016semi,wu2020connecting,deng2020cola}. We apply the following equation to update node representations:
\begin{equation}
    \mathbf{H}^{(l)}=\sigma(\tilde{\mathbf{D}}^{-1}\tilde{\mathbf{A}}\mathbf{H}^{(l-1)}\mathbf{W}^{(l-1)}),
\end{equation}

\noindent where $\tilde{\mathbf{D}}=\sum_{j}\tilde{a}_{i,j}$, $\mathbf{W}^{(l)}\in\mathbb{R}^{D\times{D}}$ is a layer-specific weight matrix, and $\mathbf{H}^{(l)}\in\mathbb{R}^{N\times{D}}$ is the node representation matrix at $l^{th}$ layer, with $\mathbf{H}^{(0)}=\mathbf{H}^{feat}$. $\sigma(\cdot)$ is the nonlinear function (e.g., exponential linear unit (ELU)).

\subsection{Prediction and Objective Function}

\noindent Due to the nonlinear characteristics of CNNs and GNNs, the scale of neural network outputs is not sensitive to the input. Moreover, the historical infection cases of each region are not purely nonlinear, especially in COVID-19 datasets, showing linear characteristics on the progression patterns of many regions, which cannot be fully handled well by neural networks \cite{zhang2003time}. To address these drawbacks, some models \cite{chakraborty2019forecasting,shih2019temporal} retain the advantages of traditional linear models and neural networks by combining a linear part to design a more robust prediction framework. Therefore, EpiGNN can optionally integrate a traditional AutoRegressive (AR) component as a linear part to obtain the linear result $\hat{\mathbf{y}}^{l}_{t+h}\in\mathbb{R}^{N}$:
\begin{equation}
    \hat{y}^{l}_{i,t+h} = \sum_{m=0}^{q-1}\mathbf{W}_m^{ar}x_{i,t-m}+b^{ar},
\end{equation}

\noindent where $q$ is the look-back window of AR, and $\mathbf{W}^{ar}\in\mathbb{R}^{q}$ is the parameters in AR component. We concatenate nodes' initial features and the output of the last layer of GCN together, and feed it to a dense layer to obtain the output:
\begin{equation}
    \hat{\mathbf{y}}_{t+h}^{n} = [\mathbf{H}^{(0)};\mathbf{H}^{(l)}] \mathbf{W}_{n} + \mathbf{b}_{n},
\end{equation}

\noindent where $[;]$ is concatenation operation, and $\mathbf{W}_{n}\in\mathbb{R}^{2D}$. The final prediction result $\hat{\mathbf{y}}_{t+h}\in\mathbb{R}^{N}$ of EpiGNN is obtained by summing $\hat{\mathbf{y}}^{l}_{t+h}$ and $\hat{\mathbf{y}}_{t+h}^{n}$:

\begin{equation}
    \hat{\mathbf{y}}_{t+h} = \hat{\mathbf{y}}_{t+h}^{l} + \hat{\mathbf{y}}_{t+h}^{n}.
\end{equation}

We employ the Mean Squared Error (MSE) to train the model by minimizing the loss. The loss function can be defined as:
\begin{equation}
    \pounds(\theta)=\left\| \mathbf{y}_{t+h}-\hat{\mathbf{y}}_{t+h} \right\|^2_2,
\end{equation}

\noindent where $\mathbf{y}_{t+h}$ is the ground truth value, and $\theta$ are all learnable parameters in EpiGNN. The pseudocode of the algorithm is described in Algorithm \ref{alg:epignn}.

\begin{algorithm}
    \caption{EpiGNN algorithm}
    \label{alg:epignn}
    \begin{algorithmic}[1]
        \Require
            Time series data $\{\mathbf{X},\mathbf{y}\}$ from multiple regions,
            geographic adjacent matrix $\mathbf{A}^{geo}$,
            external resources $\mathbf{E}$ (optional).
        \Ensure {Prediction result $\hat{\mathbf{y}}$.}
            \For{each $region$ $i$}
                \State $\mathbf{h}^{temp}_{i} \leftarrow$ Multi-Scale Convolutions($\mathbf{x}_{i:}$)
                \State $\mathbf{h}^{l}_{i} \leftarrow$ Local Transmission Risk Encoding($\mathbf{A}^{geo}_{i:}$)
                \State $\mathbf{h}^{g}_{i} \leftarrow$ Global Transmission Risk Encoding($\mathbf{h}^{temp}_{i},\mathbf{H}^{temp}$)
            \EndFor
            \For{each $region$ $pair$ $(i,j)$}
                \State $\tilde{a_{i,j}} \leftarrow$ Region-Aware Graph Learner($\mathbf{h}^{temp}_{i}$,$\mathbf{h}^{temp}_{j}$,$\mathbf{A}^{geo}$,$\mathbf{E}$)
            \EndFor
            \For{each $region$ $i$}
                \State $\mathbf{h}^{feat}_{i} \leftarrow$ $\mathbf{h}^{temp}_{i} + \mathbf{h}^{l}_{i} + \mathbf{h}^{g}_{i}$
                \State $\mathbf{h}^{(l)}_i \leftarrow$ Graph Convolution Network($\mathbf{h}^{feat}_{i},\tilde{\mathbf{A}}$)
                \State $\hat{y_i} \leftarrow$ Output($\mathbf{x}_{i:}$,$[\mathbf{h}^{feat}_{i};\mathbf{h}^{(l)}_i]$)
            \EndFor
        \State return $\hat{\mathbf{y}}$
    \end{algorithmic}
\end{algorithm}

\section{Experiments and Analysis} \label{sec:EXP}

\subsection{Experimental settings}

\subsubsection{Datasets.}
We conduct experiments on five epidemic-related datasets, three are seasonal influenza datasets and two are COVID-19 datasets. The statistics of datasets are summarized in Table \ref{tab:datasets}. All datasets have been split into training set (50\%), validation set (20\%), and test set (30\%) in chronological order.

\begin{table}[ht]
  \renewcommand{\arraystretch}{0.9} 
  \caption{Statistics of datasets, where SD is standard deviation and granularity means the frequency of epidemic surveillance records.}
  \label{tab:datasets}
  \centering
  \begin{tabular}{lccccccc}
    \hline
    Datasets & Regions & Length & Min & Max & Mean & SD & Granularity\\
    \hline
    Japan-Prefectures & 47 & 348 & 0 & 26635 & 655 & 1711 & weekly \\
    US-Regions & 10 & 785 & 0 & 16526 & 1009 & 1351 & weekly \\
    US-States & 49 & 360 & 0 & 9716 & 223 & 428 & weekly \\
    Australia-COVID & 8 & 556 & 0 & 9987 & 539 & 1532 & daily \\
    Spain-COVID & 35 & 122 & 0 & 4623 & 38 & 269 & daily \\
  \hline
\end{tabular}
\end{table}

\begin{itemize}
    \item \textbf{Japan-Prefectures} This dataset is collected from the Infectious Diseases Weekly Report (IDWR) in Japan,\footnote{https://tinyurl.com/y5dt7stm} which contains weekly influenza-like-illness (ILI) statistics from 47 prefectures from August 2012 to March 2019.
    \item \textbf{US-Regions} This dataset is the ILINet portion of the US-HHS dataset,\footnote{https://tinyurl.com/y39tog3h\label{comm:us-data}} consisting of weekly influenza activity levels for 10 Health and Human Services (HHS) regions of the U.S. mainland for the period of 2002 to 2017.
    \item \textbf{US-States} This dataset is collected from the Center for Disease Control (CDC).\textsuperscript{\ref{comm:us-data}} It contains the count of patient visits for ILI (positive cases) for each week and each state in the United States from 2010 to 2017. After removing Florida due to missing data, we keep 49 states remaining.
    \item \textbf{Australia-COVID} This dataset is publicly available at JHU-CSSE.\footnote{https://github.com/CSSEGISandData/COVID-19} We collect daily new COVID-19 confirmed cases ranging from January 27, 2020, to August 4, 2021, in Australia (including 6 states and 2 territories).
    \item \textbf{Spain-COVID} This dataset is collected by \cite{panagopoulos2021transfer}, consisting of daily COVID-19 cases for 35 administrative NUTS3 regions that were mainly affected by pandemic in Spain from February 20, 2020, to June 20, 2020. We also collect human mobility data in Spain from \textsl{Data For Good program}.\footnote{https://dataforgood.fb.com/tools/disease-prevention-maps/}
\end{itemize}

\subsubsection{Metrics.}

We adopt Root Mean Squared Error ($RMSE=\sqrt{\frac{1}{N}\sum_{i=1}^{n}(\hat{y_i}-y_i)^2}$) and Pearson's Correlation ($PCC=\frac{\sum_{i=1}^{N}(\hat{y_i}-\overline{\hat{y}})(y_i-\overline{y})}{\sqrt{\sum_{i=1}^{N}(\hat{y_i}-\overline{\hat{y}})^2}\sqrt{\sum_{i=1}^{N}(y_i-\overline{y})^2}}$) as metrics. For RMSE lower value is better, while for PCC higher value is better.



\subsubsection{Baselines.}
We compared the proposed model with the following methods:

\begin{itemize}
    \item \textbf{HA}: the historical average number of cases in observation window $T$.
    \item \textbf{AR}: the standard autoregression model.
    \item \textbf{LSTM}: the recurrent neural networks (RNN) using LSTM cell.
    \item \textbf{TPA-LSTM} \cite{shih2019temporal}: an attention-based LSTM model.
    \item \textbf{ST-GCN} \cite{yu2017spatio}: a spatial temporal graph neural network.
    \item \textbf{CNNRNN-Res} \cite{wu2018deep}: a deep learning model that combines CNN, RNN, and residual links for epidemiological prediction.
    \item \textbf{SAIFlu-Net} \cite{jung2021self}: A self-attention-based model for influenza forecasting.
    \item \textbf{Cola-GNN} \cite{deng2020cola}: a deep learning model that combines CNN, RNN and GCN for epidemic prediction.
\end{itemize}

\subsubsection{Implementation Details.}
All programs are implemented using Python 3.8.5 and PyTorch 1.9.1 with CUDA 11.1 in an Ubuntu server with an Nvidia Tesla K80 GPU. For each task we run 5 times with different random initialization. For all tasks, the batch size is set to 128, the look-back window $T$ is set to 20. The horizon $h$ is set to \{3,5,10,15\} and \{3,7,14\} for influenza and COVID-19 prediction respectively in turn. We train the model using Adam optimizer with weight decay 5e-4 and perform early stopping to avoid overfitting. We empirically choose 5 filters: \{$\mathbf{f}_{1\times{3},1}$,$\mathbf{f}_{1\times{5},1}$,$\mathbf{f}_{1\times{3},2}$,$\mathbf{f}_{1\times{5},2}$,$\mathbf{f}_{1\times{T},1}$\}. The range of hidden dimension ${F}$ is \{8,16,24,32\}, the number of CNN filters $k$ is searched from \{4,8,12,16,32\}, the dimension of pooling layer $p$ is chosen in \{1,2,3\}, the number of GCN layers $l$ is selected from 1 to 5. In COVID-19 task, the model integrates an autoregressive component as a linear part, and the window size $q$ is optimized in \{10,20\}. In Spain-COVID, we denote EpiGNN$_{exter}$ that utilizes human mobility as external resources, and the look-back window $e$ is searched from \{1,2,3\}.

\begin{table*}[htbp]
  \renewcommand{\arraystretch}{1.4} 
  \centering
  \scriptsize
  \caption{RMSE and PCC performance of different methods on three datasets with horizon = 3, 5, 10, 15. Bold face indicates the best result of each column and underlined the second-best. ${*}$ represents that the result is reported in the corresponding reference.}
  \label{tab:resultinfluenza}
  \setlength{\tabcolsep}{0.9pt}{
  \begin{tabular}{ll|cccc|cccc|cccc}
    \hline
    Dataset & & \multicolumn{4}{c|}{Japan-Prefectures} & \multicolumn{4}{c|}{US-Regions} & \multicolumn{4}{c}{US-States}\\
    \hline
    & & \multicolumn{4}{c|}{Horizon} & \multicolumn{4}{c|}{Horizon} & \multicolumn{4}{c}{Horizon}\\
    \hline
    Methods & Metric & 3 & 5 & 10 & 15 & 3 & 5 & 10 & 15 & 3 & 5 & 10 & 15\\
    \hline
    HA & RMSE & 2129 & 2180 & 2230 & 2242 & 2552 & 2653 & 2891 & 2992 & 360 & 371 & 392 & 403\\
    & PCC & 0.607 & 0.475 & 0.493 & 0.534 & 0.845 & 0.727 & 0.514 & 0.415 & 0.893 & 0.848 & 0.772 & 0.742\\
    \hline
    AR & RMSE & 1705 & 2013 & 2107 & 2042 & 757 & 997 & 1330 & 1404 & 204 & 251 & 306 & 327\\
    & PCC & 0.579 & 0.310 & 0.238 & 0.483 & 0.878 & 0.792 & 0.612 & 0.527 & 0.909 & 0.863 & 0.773 & 0.723\\
    \hline
    LSTM & RMSE & 1246 & 1335 & 1622 & 1649 & 688 & 975 & 1351 & 1477 & 180 & 213 & 276 & 307\\
    & PCC & 0.873 & 0.853 & 0.681 & 0.695 & 0.895 & 0.812 & 0.586 & 0.488 & 0.922 & 0.889 & 0.820 & 0.771\\
    \hline
    TPA-LSTM & RMSE & 1142 & 1192 & 1677 & 1579 & 761 & 950 & 1388 & 1321 & 203 & 247 & \underline{236} & 247\\
    & PCC & 0.879 & 0.868 & 0.644 & 0.724 & 0.847 & 0.814 & 0.675 & 0.627 & 0.892 & 0.833 & {0.849} & 0.844\\
    \hline
    ST-GCN & RMSE & 1115 & 1129 & 1541 & 1527 & 807 & 1038 & 1290 & 1286 & 209 & 256 & 289 & 292\\
    & PCC & 0.880 & 0.872 & 0.735 & \textbf{0.773} &  0.840 & 0.741 & 0.644 & 0.619 & 0.778 & 0.823 & 0.769 & 0.774\\
    \hline
    CNNRNN-Res & RMSE & 1550 & 1942 & 1865 & 1862 & 738 & 936 & 1233 & 1285 & 239 & 267 & 260 & 250\\
    & PCC & 0.673 & 0.380 & 0.438 & 0.467 & 0.862 & 0.782 & 0.552 & 0.485 & 0.860 & 0.822 & 0.820 & 0.847\\
    \hline
    SAIFlu-Net & RMSE & 1356 & 1430 & 1654 & 1707 & 661 & 870 & 1157 & 1215 & \underline{167} & \underline{195}  & \underline{236} & 238 \\
    & PCC & 0.765 & 0.654 & 0.585 & 0.556 & 0.885 & 0.800 & 0.674 & 0.564 & 0.930 & \underline{0.900} & \underline{0.853} & 0.852\\
    \hline
    Cola-GNN$^{*}$ & RMSE & \underline{1051} & \underline{1117} & \textbf{1372} & \underline{1475} & \underline{636} & \underline{855} & \underline{1134} & \underline{1203} &  \underline{167} & {202} & 241 & \underline{237}\\
    & PCC & \underline{0.901} & \underline{0.890} & \textbf{0.813} & \underline{0.753} & \underline{0.909} & \underline{0.835} & \underline{0.717} & \underline{0.639} & \underline{0.933} & {0.897} & 0.822 & \underline{0.856}\\
    \hline
    \hline
    EpiGNN & RMSE & \textbf{996} & \textbf{1031} & \underline{1441} & \textbf{1470} & \textbf{589} & \textbf{774} & \textbf{984} & \textbf{1061} & \textbf{160} & \textbf{186} & \textbf{220} & \textbf{236}\\
    & PCC & \textbf{0.904} & \textbf{0.908} & \underline{0.739} & \textbf{0.773} & \textbf{0.912} & \textbf{0.842} & \textbf{0.749} & \textbf{0.694} & \textbf{0.935} & \textbf{0.907} & \textbf{0.865} & \textbf{0.861}\\
    \hline
  \end{tabular}}
\end{table*}

\begin{table*}[htbp]
  \renewcommand{\arraystretch}{1.2} 
  \centering
  \caption{RMSE performance of different methods on two COVID-19 datasets with horizon = 3, 7, 14. Bold face indicates the best result of each column and underlined the second-best. - means the forecasting results are not available.}
  \setlength{\belowcaptionskip}{0pt}
  \label{tab:resultcovid}
  \setlength{\tabcolsep}{2.1mm}{
  \begin{tabular}{l|ccc|ccc}
    \hline
    Dataset & \multicolumn{3}{c|}{Spain-COVID} & \multicolumn{3}{c}{Australia-COVID}\\
    \hline
    & \multicolumn{3}{c|}{Horizon} & \multicolumn{3}{c}{Horizon}\\
    \hline
    Methods & 3 & 7 & 14 & 3 & 7 & 14\\
    \hline
    HA & 167.20 & 189.90 & 214.19 & 2948.48 & 2777.37 & 2589.61\\
    \hline
    AR & 165.07 & 179.51 & 203.13 & \underline{85.21} & 237.73 & \underline{309.03}\\
    \hline
    LSTM & 152.79 & 177.27 & \underline{184.44} & 181.97 & 315.85 & 338.34\\
    \hline
    TPA-LSTM & 150.74 & 183.52 & 227.95 & 180.14 & \underline{220.82} & 462.78\\
    \hline
    ST-GCN & 162.81 & 186.21 & 190.13 & 253.97 & 443.01 & 485.12\\
    \hline
    CNNRNN-Res & 163.75 & 208.85 & 219.65 & 210.23 & 416.90 & 488.01\\
    \hline
    SAIFlu-Net & {158.06} & {200.63} & {229.62} & 133.85 & 277.90 & 351.14 \\
    \hline
    Cola-GNN & {138.34} & {176.52} & {203.67} & 127.59 & 279.56 & 326.79 \\
    \hline
    \hline
    EpiGNN & \underline{135.54} & \underline{162.51} & {186.41} & \textbf{71.42} & \textbf{153.07} & \textbf{287.90}\\
    \hline
    EpiGNN$_{exter}$ & \textbf{129.90} & \textbf{145.33} & \textbf{178.73} & - & - & - \\
    \hline
  \end{tabular}}
\end{table*}

\begin{table*}[htbp]
  \renewcommand{\arraystretch}{1.05} 
  \centering
  \scriptsize
  \caption{Runtime (s) and model size (K) comparison on three influenza datasets when horizon=5. Runtime is the time spent on a single GPU per epoch.}
  \setlength{\belowcaptionskip}{0pt}
  \label{tab:resultoftimeandparam}
  \setlength{\tabcolsep}{2mm}{
  \begin{tabular}{lccccccc}
    \hline
    \multirow{2.5}{*}{Dataset ($h=5$)} & \multicolumn{2}{c}{Japan-Prefectures} & \multicolumn{2}{c}{US-Regions} & \multicolumn{2}{c}{US-States}\\
    \cmidrule(r){2-3} \cmidrule(r){4-5} \cmidrule(r){6-7}
    & Runtime & Params. & Runtime & Params. & Runtime & Params.\\
    \hline
    ST-GCN & 0.18 & 27K & 0.16 & 26K & 0.18 & 27K\\
    CNNRNN-Res & 0.05 & 13K & 0.04 & 5K & 0.06 & 14K \\
    SAIFlu-Net & 0.15 & 35K & 0.10 & 26K & 0.14 & 32K\\
    Cola-GNN & 0.14 & 9K & 0.13 & 7K & 0.15 & 9K \\
    EpiGNN (ours) & 0.10 & 11K & 0.14 & 9K & 0.07 & 12K \\
    \hline
  \end{tabular}}
\end{table*}

\subsection{Prediction Performance}
\noindent We evaluate each model in short-term (horizon $<$ 10) and long-term (horizon $\geq$ 10 ) settings. The experimental results on influenza datasets and COVID-19 datasets are shown in Table \ref{tab:resultinfluenza} and Table \ref{tab:resultcovid} respectively. There is an overall trend that the prediction accuracy drops as the prediction horizon increases because the larger the horizon, the harder the problem. The large difference in RMSE across different datasets is due to the scale and variance of the datasets. 

We observe that EpiGNN outperforms other models on most tasks. EpiGNN achieves 5.6\% and 13.4\% lower RMSE than the best baselines in the influenza prediction task and COVID-19 prediction task respectively. In influenza prediction tasks, most deep learning-based models perform better than statistical models (i.e., HA/AR) since they make effort to deal with nonlinear characteristics and complex patterns behind time series. We also notice that statistical model AR is competitive on COVID-19 prediction tasks, especially on Australia-COVID dataset. This could be because of the strong seasonal effects of influenza datasets, which is obviously not the situation in the COVID-19 historical statistics. During COVID-19 period, due to government interventions (e.g., stay-at-home orders, lockdown), the epidemic situations of regions show significant differences. It turns out that a simple linear aggregation over the past case numbers can achieve relatively good performance. EpiGNN also achieves the best performance in COVID-19 datasets attributed to the integration of a linear model. In Spain-COVID, we conduct EpiGNN$_{exter}$ which considers human mobility data as external information in E.q \ref{eq:external} to distill the correlations between regions by providing more practical evidence. The results exhibit that EpiGNN$_{exter}$ is better than EpiGNN, pointing out that external information is helpful for capturing correlations between regions. Table \ref{tab:resultoftimeandparam} shows the runtimes and number of parameters for each model on influenza datasets. EpiGNN has no obvious adverse effect on training efficiency and well controls the model size to prevent overfitting.

\subsection{Ablation Study}

\begin{figure}[htbp]
\centering
\includegraphics[width=0.72\textwidth]{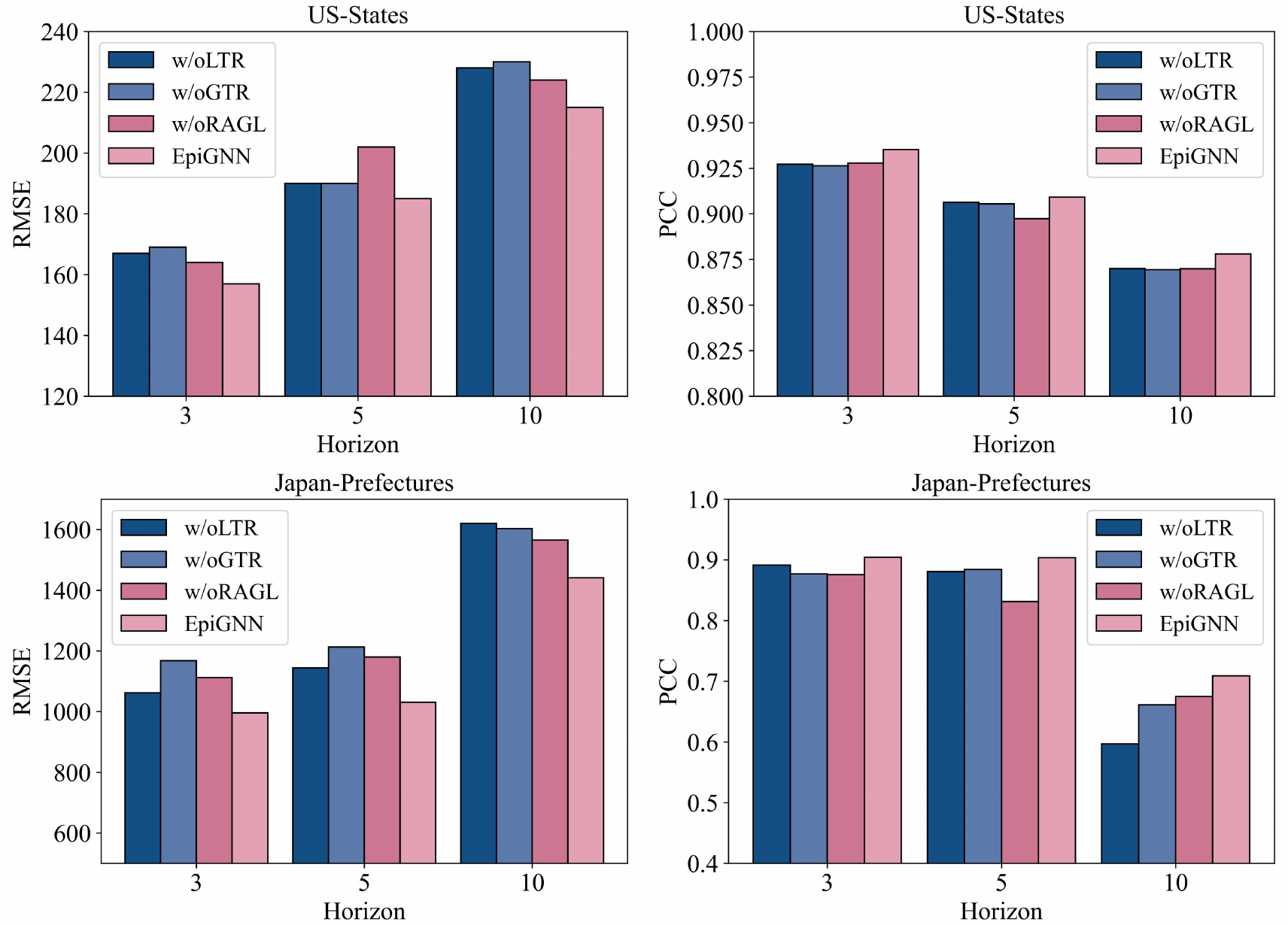}
\caption{Results of ablation studies on US-States (top) and Japan-Prefectures (bottom) datasets. For RMSE lower value is better, while for PCC higher value is better.} \label{fig:ablation}
\end{figure}

%

\begin{itemize}
    \item \textbf{w/oLTR} stands for EpiGNN without local transmission risk encoding.
    \item \textbf{w/oGTR} represents EpiGNN without global transmission risk encoding.
    \item \textbf{w/oRAGL} indicates EpiGNN using self-attention \cite{vaswani2017attention} to capture dependencies between regions instead of Region-Aware Graph Learner (i.e., applying $\tilde{\mathbf{A}}=\text{softmax}((\mathbf{H}^{feat}\mathbf{W}_{1})(\mathbf{H}^{feat}\mathbf{W}_{2})^{T})$).
\end{itemize}

We perform ablation studies on Japan-Prefectures and US-Regions datasets, and the results measured using RMSE and PCC are shown in Fig. \ref{fig:ablation}. We quantitatively show that the complete EpiGNN can yield the most stable and optimal performance compared to other incomplete models. Compared with using self-attention, RAGL can bring performance gains. The fact can be attributed that RAGL well utilizes spatial and temporal information, which affirms the importance of designing a suitable approach to explore the correlations between regions. In addition, since the captured dependencies are not fully bidirectional, it helps GCN to focus on potentially related regions to overcome the oversmoothing phenomenon \cite{li2018deeper} and avoid noise accumulation. We also notice that both w/oLTR and w/oGTR cause performance drops, which indicates the positive impacts of transmission risk encodings, and exhibits the effectiveness of modeling transmission risks because they emphasize the spatial effects of regions and provide interpretable evidence on risky areas.

\begin{figure}
\centering
\includegraphics[width=0.9\textwidth]{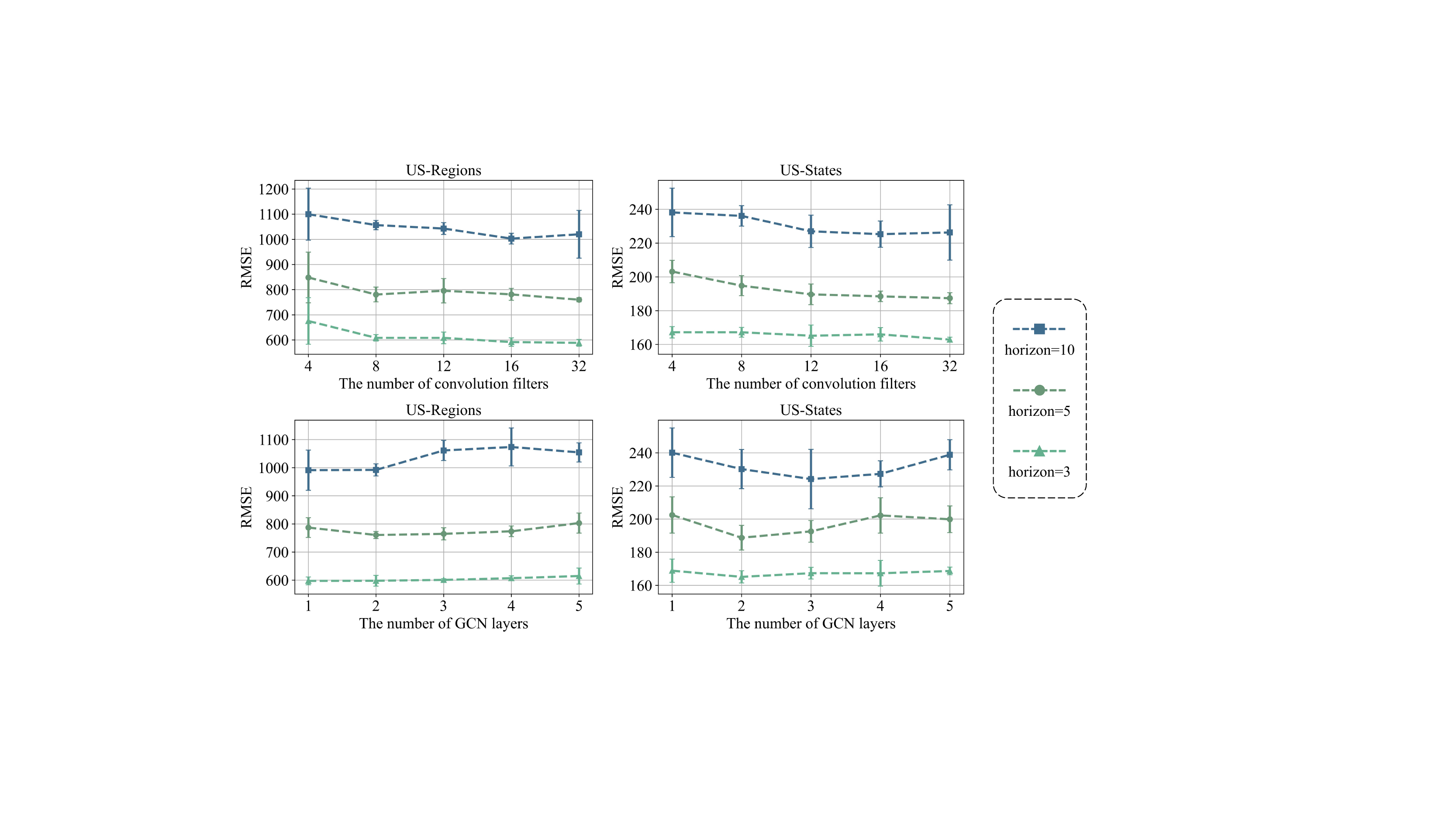}
\caption{Parameters analysis results of convolution filter number $k$ (top) and GCN layer number $l$ (bottom) on US-Regions (left) and US-States (right) datasets.} \label{fig:parameter}
\end{figure}

\subsection{Parameters Analysis}


\subsubsection{Number of convolution filters.}

Different convolution filters learn different features behind data. We evaluate $k$ in range \{4,8,12,16,32\}, and the results are shown in Fig. \ref{fig:parameter}. Smaller $k$ results in poor prediction performance due to limited representation ability. As $k$ increases, there are more learnable parameters in model and could bring performance gain to a certain extent. We recommend selecting $k$=12 to achieve a balance between accuracy and computation.

\subsubsection{Number of GCN layers.}

More GCN layers stacked tend to aggregate nodes' features from wider neighborhood ranges. We vary the number of GCN layers from 1 to 5, and the results are shown in Fig. \ref{fig:parameter}. We observe that smaller $l$ can reach better performance. However, performance drops when $l$ increases reveals that integrating information from irrelevant/weakly-related nodes may result in oversmoothing \cite{li2018deeper} or bring noises, which will undermine the performance.

\subsection{Visualization}

\begin{figure}
\includegraphics[width=\textwidth]{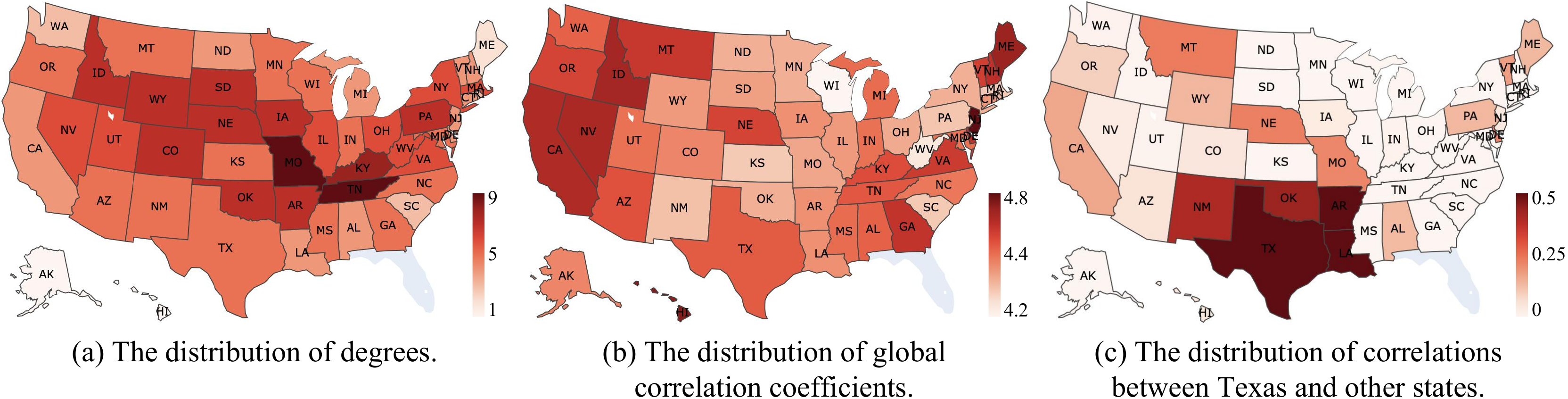}
\caption{Visualization of intermediate results.} \label{fig:usa_example}
\end{figure}

\noindent We visualize an example with window=(2016/46$^{th}$-2017/13$^{th}$) and horizon=5 (week) in US-States dataset, meanwhile, we also provide potential risky regions. Fig. \ref{fig:usa_example}(a) is the distribution of {degrees} in the United States. We notice that the more central/larger region, the greater the degree. Fig. \ref{fig:usa_example}(b) is the distribution of global correlation coefficients. Compared with Fig. \ref{fig:usa_example}(a), it can be seen that some states (e.g., CA) that are not in the center have high global correlation coefficients. Texas (TX) is the largest and second-most populous state in the U.S. which has a relatively high {degree} and global correlation coefficient in this case study. We show how Texas is related to other states as drawn in Fig. \ref{fig:usa_example}(c). In Fig. \ref{fig:usa_example}(c), Texas does not have dependencies with all states. Nevertheless, Texas has relatively significant dependencies with its adjacent regions and also has relationships with some non-adjacent regions.

\begin{figure}
\centering
\includegraphics[width=0.99\textwidth]{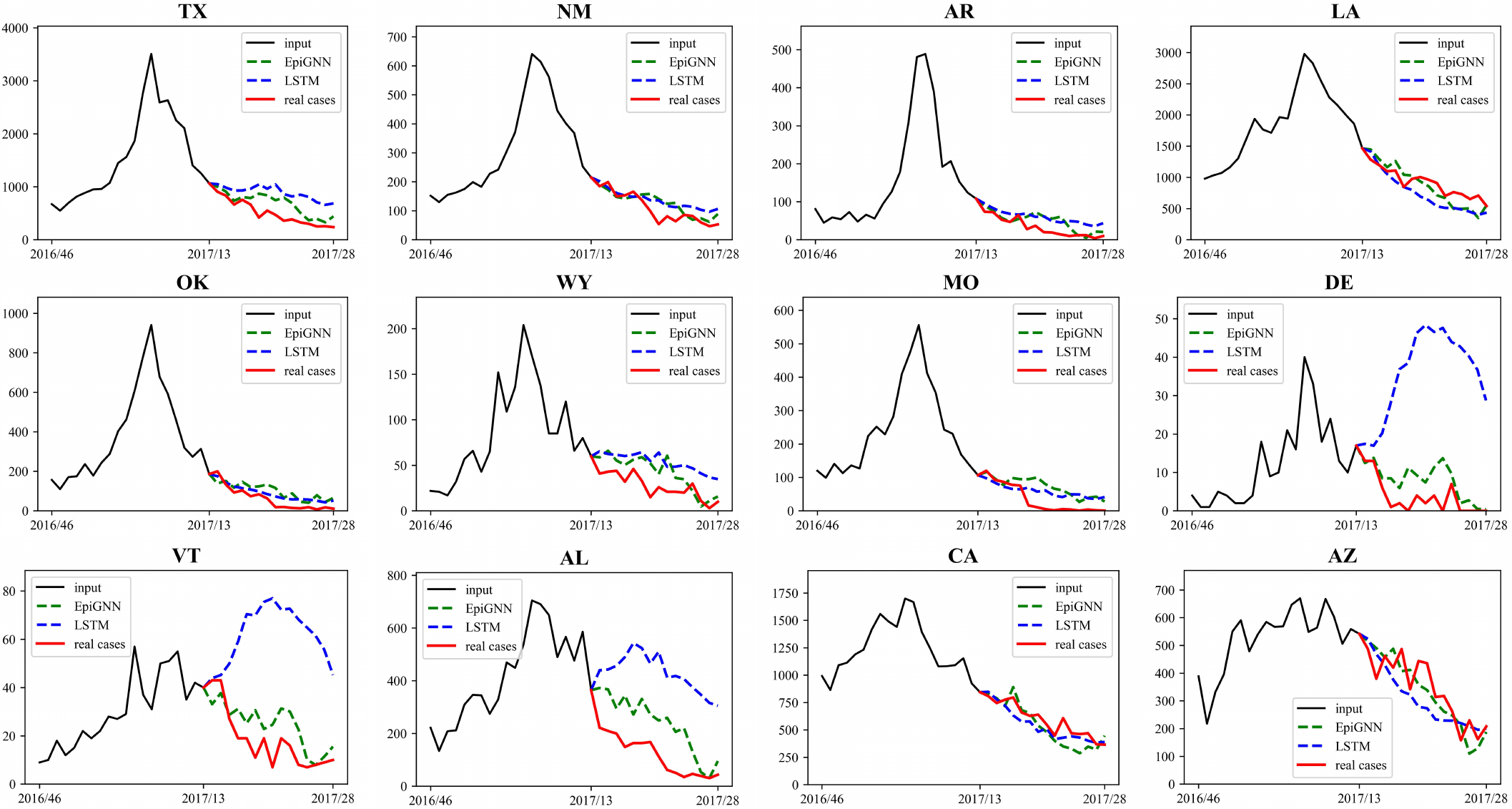}
\caption{Predicted curve of EpiGNN (green) and LSTM (blue) for selected states.} \label{fig:usa_result}
\end{figure}

We visualize the predicted curve of EpiGNN and LSTM in Fig. \ref{fig:usa_result}. Compared with LSTM, we observe that EpiGNN fits the ground truth better, and some trends of fluctuation are also predicted better (e.g., WY/DE/VT), while LSTM yields quite inaccurate predictions in some states. We notice that there are similar progression patterns between TX and its adjacent states (e.g., NM/AR/LA), which indicates that local correlations between geographically adjacent regions may be very strong. The correlations drawn in Fig. \ref{fig:usa_example}(c) also show that adjacent regions are strongly related, which is consistent with the existing finding \cite{mcmahon2022spatial}.


\section{Conclusions} \label{sec:CON}

\noindent In this paper, we develop EpiGNN, a novel model for epidemic prediction. In this model, we design a transmission risk encoding module to characterize local and global spatial effects of each region. Meanwhile, we propose a Region-Aware Graph Learner that takes transmission risk, geographical dependencies, and temporal information into account to better explore spatial-temporal dependencies. Experimental results show the effectiveness and efficiency of our method on five epidemic-related datasets. As for future work, we will devote to better predict by considering the time decay effects of spatial transmission. \newline

\noindent\textbf{Acknowledgment.} This work is supported by the Key R\&D Program of Guangdong Province No.2019B010136003 and the National Natural Science Foundation of China No. 62172428, 61732004, 61732022.

%
%

%
%
%

\bibliographystyle{splncs04}

%
\end{document}